# Adsorption, Segregation and Magnetization of a Single Mn Adatom on the GaAs (110) Surface

J. X. Cao[1,2], X. G. Gong[1,3] and R. Q. Wu[1,4]

1. Interdisciplinary Center of Theoretical Studies, Chinese Academy of Sciences, 100080 Beijing, China
2. Department of Physics, Xiangtan University, 411105, Xiangtan, China
3. Surface Science Laboratory (State Key) and Department of Physics, Fudan University, 200433 Shanghai, China
4. Department of Physics and Astronomy, University of California, Irvine, California 92697, USA

**Abstract:**

*Density functional calculations with a large unit cell have been conducted to investigate adsorption, segregation and magnetization of Mn monomer on GaAs(110). The Mn adatom is rather mobile along the trench on GaAs(110), with an energy barrier of 0.56 eV. The energy barrier for segregation across the trenches is nevertheless very high, 1.67 eV. The plots of density of states display a wide gap in the majority spin channel, but show plenty of metal-induced gap states in the minority spin channel. The Mn atoms might be "invisible" in scanning tunneling microscope (STM) images taken with small biases, due to the directional p-d hybridization. For example, one will more likely see two bright spots on Mn/GaAs(110), despite the fact that there is only one Mn adatom in the system.*

Diluted magnetic semiconductors (DMS) offer superior compatibilities to base semiconductors and are highly promising as robust spin injectors for spintronics manipulations[1,2,3,4]. Despite extensive explorations, however, exploitation of DMS seems still severely hindered by several major complexities such as lack of understanding in the magnetization and uncontainable growth processes. For (Ga,Mn)As, a prototype DMS that has attracted extensive attention recently[5], it was found that that low temperature annealing procedures are necessary to increase the as-grown $T_c$ to 110-160 K[6,7,8]. The diffusion of interstitial Mn atoms toward the surface region is found to play a key role for the increases in $T_c$. On the other hand, Mn atoms in the surface region can be converted to substitutional dopants, as revealed in recent joint studies through density functional calculations and cross-sectional scanning tunneling microscope (STM)[9,10] measurements. Understanding in adsorption and segregation of magnetic impurities on semiconductor surfaces hence becomes indispensable for further comprehensive studies of DMS, and moreover for the design and fabrication of excellent magnetic wires, dots and interfaces. So far, extensive explorations have been conducted for Mn/GaAs(001), chiefly due to the standard set by the chip industry. Yet, on the contrary, publications for Mn/GaAs(110) can be hardly found in the literature. This motivated us to carry out the present theoretical investigations for the adsorption, segregation and magnetization of Mn/GaAs(001), starting from cases with a single adsorbate.

In the bulk zinc-blende semiconductors, there are three high symmetric interstitial positions, namely, the two tetrahedral sites which is surround by four nearest neighbors cations (the $T_{Ga}$ site) or anions (the $T_{As}$ site) and one hexagonal site which is surround by six nearest neighbors (three cations and three anions). Edmonds et al[8] estimated that the energy barrier governing the diffusion of interstitial Mn in the bulk (Ga,Mn)As is 0.7±0.1 eV. The energy barriers have also been calculated by several groups and results are still somewhat controversial. Through the full potential linearized augmented plane wave (FLAPW) calculation, Masek et al[11] found that the formation energies are very close for Mn in either the $T_{As}$ or $T_{Ga}$ site. The hexagonal interstitial site, however, is higher in energy by 0.52 eV. In contrast, results obtained from plane wave calculations indicate that $T_{Ga}$ sites is about 0.35 eV higher in energy for a single interstitial Mn and the energy barrier for its diffusion is 0.8 eV[8]. Results of our own calculations are close to Masek's data if the positions of Ga and As atoms are frozen. On the contrary, the total energies for Mn at all the three sites are almost the same when the lattice size and atomic positions of GaAs are fully relaxed. In this paper, we report that Mn monomer migrates along the trench on the GaAs(110) surface with an energy barrier of 0.56 eV. $T_{Ga}$ is the most stable adsorption site while $T_{As}$ is a metastable adsorption site. The magnetic moment of Mn/GaAs(110) is 3.0 $\mu_B$. Importantly, Mn induces pronounced resonance features in the band gap in the minority spin channel. The wave functions of these gap states are very directional, and unusual STM images are predicted.

To determine the potential energy surface (PES), we calculated the binding energies of Mn on seven different adsorption sites as shown in Fig.1. Among these sites, III, IV and VI correspond to the hexagonal position while V and VII are close to the $T_{Ga}$ and $T_{As}$ sites in the bulk GaAs, respectively. The projected augmented wave (PAW) approach and the generalized gradient approximation (GGA), implemented in the VASP code, were used in the present calculations. We modeled the GaAs (110) surface with a slab geometry that contains five GaAs layers and a vacuum of 10.0 Å thick in the between. To minimize the remaining size effect due to the use of finite slab, the bottom surface is passivated with a layer of H. In the lateral plane, we used a 3×3 super cell so as to mimic the single adsorbate environment. Energy cutoff of 300 eV was chosen for the plane wave basis expansions. 4×4 special *k*-mesh points in the Brillouin-zone were used to evaluate integrations in the reciprocal space. The vertical position of the Mn adatom is fully relaxed for a given (x,y) position. The positions of Ga and As atoms in the topmost three layers are also fully optimized, whereas those in the two bottommost layers are frozen at their bulk positions. The criterion for structural optimization requires the calculated force on each unconstrained atom smaller than 0.01 eV/Å. Results of multilayer relaxations for the clean GaAs(110) obtained from test calculations agree very well with data in the literature, indicating the validity of the present treatment. We found that a large search range should be given to the z-coordinate of Mn since local minima are frequently involved when the substrate relaxes simultaneously.

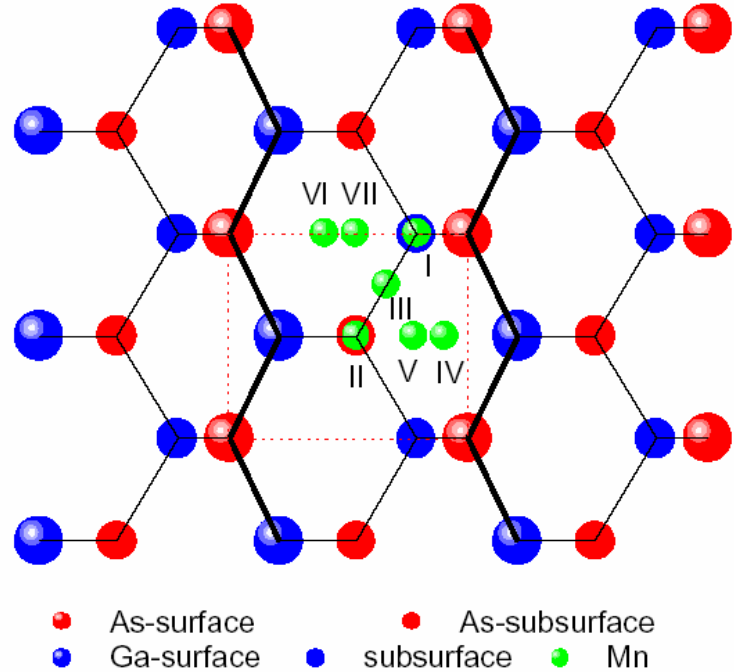

*Fig.1, Sketch of the p(3×3) unit cell and adsorption configurations for the simulations of Mn/GaAs(110). The dashed rectangle gives the primitive 2D unit cell for the clean GaAs(110) surface.*

The calculated two-dimensional PES for Mn/GaAs(110) displays a strong spatial anisotropy in Fig. 2. The red line indicates that the apparent segregation path for Mn is along the trench on the GaAs(110) surface. Quantitatively, the total energy increases very steeply away from its minimum at a point close to the $T_{Ga}$ site. To pass through the site III toward the metastable $T_{As}$ site, the Mn adatom needs to overcome an energy barrier of 0.56 eV. Nevertheless, the energy barrier for the reverse trip is very shallow, only 0.2 eV. In contrast, there is no obvious channel for Mn to segregate across the trenches and the smallest energy barrier for this segregation is as high as 1.67 eV. Therefore, Mn migration on GaAs(110) is one-dimensional at ambient temperature, only along the trench. Although it is expected that the Mn adatom resides mostly on the $T_{Ga}$ site, the small energy barriers make it rather mobile. So far, there is no direct measurement for the segregation barriers on Mn/GaAs(110). Density functional calculations done by Ishii et al found similar size of energy barriers, 0.57 eV, for migrations of Ga and As along the trench[12]. The barrier height estimated for Mn impurities in bulk (Ga,Mn)As is somewhat larger, 0.7±0.1 eV, but this value might be affected by the attraction between the interstitial and substitutional Mn[8].

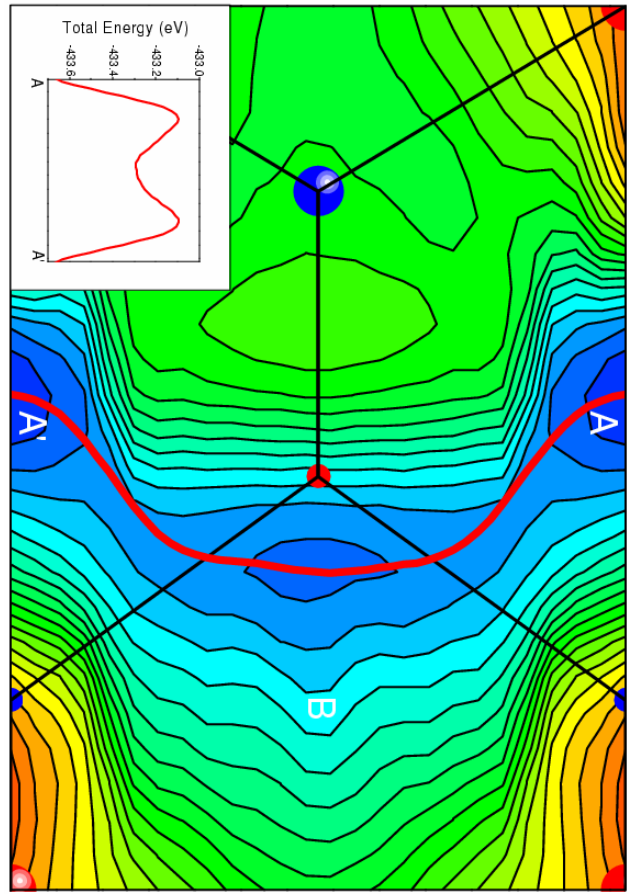

*Fig. 2. Potential-energy surface for Mn diffusion on GaAs(110). Cold colors indicate higher binding energies whereas warm colors represent lower binding energies. Balls and straight black lines give the positions of the surface and subsurface Ga and As atoms and the bonds between them. The red line shows the minimum-energy segregation path and the inset gives the total energy versus the position of Mn along the path.*

Table I presents the optimized z-positions of the Mn adatom, the nearest Mn-Ga and Mn-As interatomic distances, the total magnetic moments, and the local magnetic moments along with the numbers of electrons integrated in the Mn muffin-tin region ($r_{Mn}$=1.17 Å). Strikingly, the value of z remains negative except on the unstable sites, I and II (cf. Figs. 1 and 2) --- indicating that Mn stays below the surface As layer when it migrates in the trench. The Mn appears to be positively charged on GaAs(110) and show an attractive (repulsive) tendency toward As (Ga). This is manifested by the fact that $d_{Mn-As}$ is smaller than $d_{Mn-Ga}$ by 0.17 Å on the $T_{As}$ (or V) site whereas they are almost the same on the $T_{Ga}$ (or VII) site. In addition, Mn stays far above the surface on site I to avoid a close contact with the Ga cation underneath. As a matter of fact, the amount of electron in Mn varies slightly from place to place. It possesses fewer electrons when surrounded by Ga atoms, and vise versa. Strikingly, we found only small difference in $N_{MT}$ for Mn in the surface and bulk environments (< 0.15 e). Therefore, they are not negatively charged, from the point view of $N_{MT}$, even though interstitial Mn impurities are perceived as double donors in the bulk (Ga,Mn)As. Furthermore, the total magnetic moment of Mn/GaAs(110) is 3.0 $\mu_B$ in most geometry. The contribution in the Mn muffin-tin region, however, fluctuates in a wide range. For example, $M_{MT}$ is 3.57 $\mu_B$ on the $T_{Ga}$ site but it becomes only 3.06 $\mu_B$ on the $T_{As}$ site. Obviously, the Mn-induced magnetic moments on Ga and As atoms also change drastically for different adsorption sites. For bulk (Ga,Mn)As[13], the magnetic moment of interstitial Mn (at the center of Ga-tetrahedron) is 3.1 $\mu_B$ through our previous FLAPW calculations.

*Table I. The z-position of Mn (measured from the topmost As layer of the relaxed clean GaAs(110) surface), the nearest Mn-Ga and Mn-As interatomic distances, the total and local magnetic moment as well as the number of electrons integrated in the Mn Muffin-tin sphere for different adsorption configurations.*

| Adsorption site | z (Å) | $d_{Mn-Ga}$ (Å) | $d_{Mn-As}$ (Å) | $N_{MT}$ (electron) | M ($\mu_B$) | $M_{MT}$ ($\mu_B$) |
|---|---|---|---|---|---|---|
| I | 2.00 | | 2.58 | 4.99 | 5.00 | 4.33 |
| II | 0.39 | 2.36 | 2.62 | 5.05 | 5.00 | 4.08 |
| III | -0.29 | 2.45 | 2.47 | 5.19 | 3.00 | 3.41 |
| IV | -1.03 | 2.47 | 2.51 | 5.23 | 3.00 | 3.16 |
| V | -0.66 | 2.50 | 2.33 | 5.28 | 3.00 | 3.06 |
| VI | -0.73 | 2.44 | 2.46 | 5.21 | 3.00 | 3.27 |
| VII | -0.24 | 2.57 | 2.59 | 5.12 | 3.00 | 3.57 |

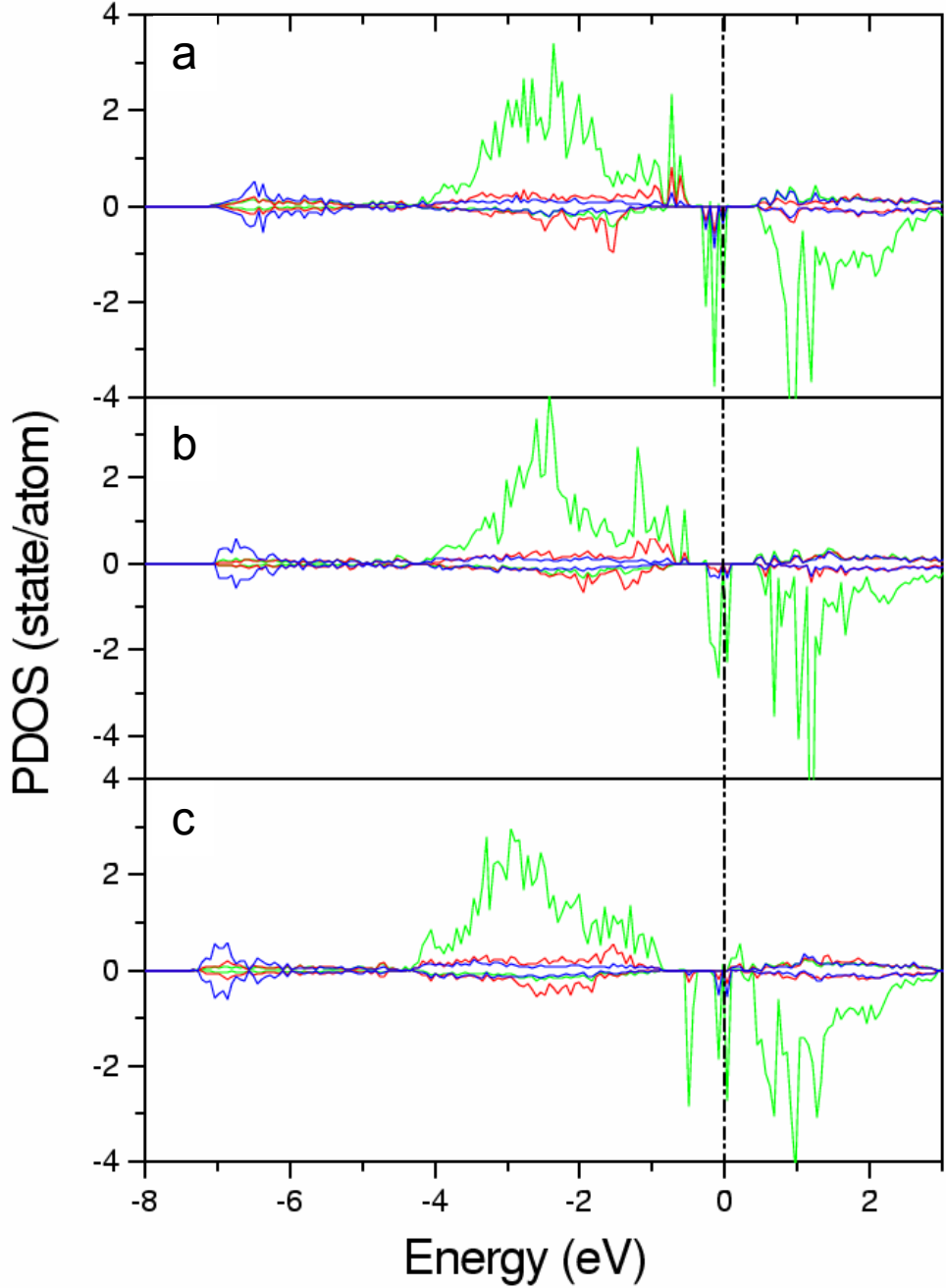

*Fig. 3. Atom-projected density of states for Mn(green lines) and its nearest Ga(blue lines) and As (red lines) neighbors. Results in panels (a), (b) and (c) are for $T_{Ga}$, $T_{As}$ and hexagonal (site III) configurations, respectively. States in the majority and minority spins are separated with positive and negative signs. Zero energy in the horizontal axis indicates the position of the Fermi level.*

The integer total magnetic moments listed in Table I are typically associated with the half-metallic feature. As seen from the plots of the projected density of states (PDOS) in

Fig. 3, large exchange splittings can be found for the Mn d-bands. The majority spin part is fully occupied whereas the minority spin bands remain mostly empty. The Mn d-band width is large, 3 eV, a result which suggests strong p-d hybridization between Mn and the surrounding Ga and As atoms. Remarkably, both spin channels have band gaps, but the Mn-induced resonances around $E_F$ can be found only in the minority spin channel. These gap states stem from the p-d hybridization between Mn and their Ga and As neighbors. In Fig. 3, it is clear that they distribute mainly in Mn, but the weights in Ga and As are also sizable. The strongest p-d intermixing occurs to the $T_{Ga}$ site, the most stable adsorption configuration. Wave functions of these gap states are found to be strongly directional for the $T_{Ga}$ case, manifested by lobs between Mn and Ga. By contrast, Mn-As bonding is more isotropic for the $T_{As}$ adsorption case.

Experimentally, it is possible to observe these gap states through STM under small bias. We present their charge densities (from states within ±0.25 eV) in Fig. 4 for both $T_{As}$ and $T_{Ga}$ configurations. Interestingly, no electron cloud can be found on the top of the Mn atom in the plane 4 Å above Mn, because of strong localization of Mn-3d states and the anisotropic features of Mn-Ga bonds mentioned above. For the $T_{Ga}$ adsorption case, surprisingly, two distinguished lobs, 4.0 Å apart, can be seen above the middle points of the bonds between the surface Ga and subsurface As. One will thus expect two bright spots from Mn/GaAs(110) in small bias STM images, even though there is only one Mn adatom in the system. For the metastable $T_{As}$ configuration, possibly also observable at low temperature, a blur cigar shape image is expected according to the present calculations. In both cases, the Mn atom appears to be "invisible" through STM under small biases. One hence needs to be very careful in interpreting STM images taken for metal impurities on semiconductor surfaces. Since the gap states only exist in the minority spin channel, they also carry spin signal and should play an important role in mediating the magnetic ordering.

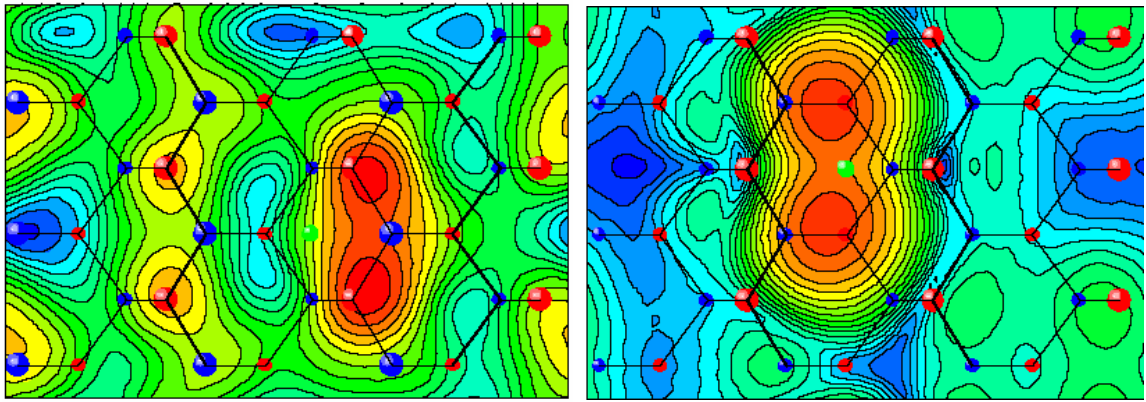

*Fig. 4. The energy-sliced charge densities plotted in a horizontal plane which is 4 Å above the Mn adatom. States only within -0.25 eV to +0.25 eV around $E_F$ contribute to the plots. Panels (a) and (b) represent results for $T_{As}$ and $T_{Ga}$ configurations, respectively.*

In conclusion, density functional calculations revealed interesting segregation behaviors for Mn monomer on GaAs(110) . It is shown that Mn prefer the $T_{Ga}$ site but remains rather mobile along the trench, a result which is very important for understanding the growth dynamics. Peculiar electronic and magnetic properties, especially the extraordinary STM patterns predicted here, deserve experimental verifications. Understanding in interaction between Mn and GaAs from different angles is necessary for the establishment of more comprehensive pictures.

This work was supported by National Natural Science Foundation of China (No. 10447132), China Postdoctoral Science Foundation(No. 2004036085), K. C. Wong Education Foundation, and partly by the Science and Technology Foundation for Younger of Hunan Province. The computation was performed on the HP-SC45 Sigma-X parallel computer of ITP and ICTS, CAS. XGG is supported by National Natural Science Fundation of China, the special funds for major state basic research and CAS projects. RW is indebted to Prof. A. Yazdani for stimulating discussions.